\documentclass[journal=jacsat,manuscript=article,layout=twocolumn,articletitle=true]{achemso}

\setkeys{acs}{usetitle = true}
\usepackage[version=3]{mhchem} % Formula subscripts using \ce{}

\usepackage{caption}
\usepackage{graphicx}
\usepackage{epstopdf}
\usepackage{dcolumn}
\usepackage{amsmath}
\usepackage{epsfig}
\usepackage{indentfirst}
\usepackage{psfrag}
\usepackage{amssymb}
\usepackage{xcolor}
\usepackage{float}
\usepackage{newfloat}
\usepackage{bm}% bold math
\usepackage{soul}

\newcommand{\np}[1]{\textcolor{black}{#1}}
\newcommand{\agln}[1]{\textcolor{teal}{#1}}

\definecolor{mypink3}{cmyk}{0.98, 0, 0.36, 0.80}

\newcommand{\onlinecite}[1]{\hspace{-1 ex} \nocite{#1}\citenum{#1}} 

\def\didv{dI/dV} 
\def\vs{\mbox{V$_{\textrm{s}}$}}
\def\Vs{\mbox{V$_{\textrm{s}}$}}
\def\It{\mbox{I$_{\textrm{t}}$}}

\author{Eduard Carbonell-Sanrom\`{a}}  
   \affiliation{CIC nanoGUNE,  20018 Donostia-San Sebasti\'an, Spain}

\author{Aran Garcia-Lekue}
   \affiliation{Donostia International Physics Center, 20018 Donostia-San Sebasti\'an, Spain}
   \alsoaffiliation{Ikerbasque, Basque Foundation for Science, 48013 Bilbao, Spain}
   \email{wmbgalea@ehu.eus}

\author{Martina Corso}
   \affiliation{Centro de F{\'{\i}}sica de Materiales CFM/MPC (CSIC-UPV/EHU), 20018 Donostia-San Sebasti\'an, Spain}
   \alsoaffiliation{Donostia International Physics Center, 20018 Donostia-San Sebasti\'an, Spain}
   \email{martina.corso@ehu.eus}

\author{Guillaume Vasseur}
\affiliation{Centro de F{\'{\i}}sica de Materiales
	CFM/MPC (CSIC-UPV/EHU),  20018 Donostia-San Sebasti\'an, Spain}

\author{Pedro Brandimarte}
   \affiliation{Donostia International Physics Center, 20018 Donostia-San Sebasti\'an, Spain}

\author{Jorge Lobo-Checa}
   \affiliation{Instituto de Ciencia de Materiales de Arag\'on (CSIC-UZ), E-50009 Zaragoza, Spain}
   \alsoaffiliation{Dpto. F\'{\i}sica de la Materia Condensada, Univ. de Zaragoza, E-50009 Zaragoza, Spain}   
   
\author{Dimas G. de Oteyza}
   \alsoaffiliation{Donostia International Physics Center, 20018 Donostia-San Sebasti\'an, Spain}
   \alsoaffiliation{Ikerbasque, Basque Foundation for Science, 48013 Bilbao, Spain}

\author{Jingcheng Li}
   \affiliation{CIC nanoGUNE,  20018 Donostia-San Sebasti\'an, Spain}

\author{Shigeki Kawai}
   \affiliation{International Center for Materials Nanoarchitectonics, National Institute for Materials Science, 1-1, Namiki, Tsukuba, Ibaraki 305-0044, Japan.}

\author{Shohei Saito}
   \affiliation{Graduate School of Science, Kyoto University, Kyoto 606-8502, Japan}

\author{Shigehiro Yamaguchi}
   \affiliation{Graduate School of Science, Nagoya University, Nagoya 464-8602, Japan}

\author{J. Enrique Ortega}
   \affiliation{Centro de F{\'{\i}}sica de Materiales
	CFM/MPC (CSIC-UPV/EHU),  20018 Donostia-San Sebasti\'an, Spain}
\author{Daniel S\'anchez-Portal}
   \affiliation{Centro de F{\'{\i}}sica de Materiales
	CFM/MPC (CSIC-UPV/EHU),  20018 Donostia-San Sebasti\'an, Spain}
\author{Jose Ignacio Pascual}
   \affiliation{CIC nanoGUNE,  20018 Donostia-San Sebasti\'an, Spain}
   \alsoaffiliation{Ikerbasque, Basque Foundation for Science, 48013 Bilbao, Spain}
   \email{ji.pascual@nanogune.eu}

\title{Electronic Properties of Substitutionally Boron-doped Graphene Nanoribbons \\ on a  Au(111) Surface}

\begin{document}
\date{\today}
  
\begin{abstract}
{\bf High quality graphene nanoribbons (GNRs) grown by on-surface synthesis strategies with atomic precision can be controllably doped by inserting heteroatoms or chemical groups in the molecular precursors. Here, we study the electronic structure of armchair GNRs substitutionally doped with di-boron moieties at the center, through a combination of scanning tunneling spectroscopy, angle-resolved photoemission, and density functional theory simulations. Boron atoms appear with a small displacement towards the surface signaling their stronger interaction with the metal. We find two boron-rich flat bands emerging as impurity states inside the GNR band gap,  one of them particularly broadened after its hybridization with the gold surface states.  In addition, the boron atoms  
shift the conduction and valence bands of the pristine GNR away from the gap edge, and leave unaffected the bands above and below, which become the new  
frontier bands and have negligible boron character. This is due to the selective mixing of boron states with GNR bands according to their symmetry. \np{Our results depict that the GNRs band structure can be tuned by modifying the  
separation between di-boron moieties.}  
}
\end{abstract}
  
%\linenumbers

\section{Introduction}

Graphene nanoribbons (GNRs) have emerged as a low dimensional platform for nanoelectronics with  promising applications due to the predictive control on their functionality \cite{Celis2016}. Most GNRs are predicted to be one-dimensional semiconductors, with a band-gap that depends on their width  and orientation \cite{Yang2007,Wakabayashi2010}. However, due to their low dimensions, their electronic properties are extremely sensitive to atomic-scale defects and inhomogeneities appearing during their fabrication. On-surface chemistry strategies allow producing atomically precise nanoribbons with perfectly well-defined widths and edges, using organic precursors and directing their reaction over a metal surface \cite{Cai2010,Tarliz2016rev,corso_rev2018}. This method succeeded in fabricating nanoribbons with controlled size
\cite{Cai2010,zhang_-surface_2015,chen_ACSNano2013,talirz_-surface_2017,abdurakhmanova_Carbon2014,MerinoACS17}
and orientation \cite{de_oteyza_substrate-independent_2016,ruffieux_-surface_2016}, in doping  them with chemical groups \cite{nguyen_bottom-up_2016, kawai_atomically_2015,cloke_site-specific_2015,carbonell-sanroma_quantum_2017,Carbonell2017CN,Durr_JACS2018,Kawai2018}, as well as in producing hybrid structures with enhanced functionality \cite{cai_graphene_2014,chen_molecular_2015,Li2018}.     

Here we study the electronic structure of 7 atoms wide armchair GNRs (7-AGNRs) substitutionally doped with di-boron moieties at the center of the carbon backbone (2B-7-AGNRs, Scheme \ref{precursor}), adsorbed on a Au(111) surface.  In our study, we combine  scanning tunneling spectroscopy (STS), angle-resolved photoemission (ARPES), and density functional theory (DFT) simulations to unveil the effect of inserting boron atoms in the electronic structure of the pristine ribbon. The synthesis of these ribbons was previously reported in Refs.\ \onlinecite{kawai_atomically_2015,cloke_site-specific_2015}, \np{and a recent report  proposed that their bands should exhibit a different topological configuration than bare 7-AGNRs \cite{Cao2017a}. } We find that the di-boron impurities modify the electronic structure of the 7-AGNR in two ways. First, the boron states interact with both valence and conduction bands (VB and CB) of the pristine 7-AGNR,   while the VB-1 and CB+1 bands remain unaffected. This causes a band-inversion of the frontier bands. 
Second, two impurity (boron-rich) flat bands appear  above and below the Fermi level (E$_{\rm{F}}$). The unoccupied one, already predicted in Refs.\ \onlinecite{kawai_atomically_2015,cloke_site-specific_2015}, is found here as a broad band, strongly mixed with Au(111) states. Both impurity bands lie well inside the 7-AGNR band gap and their inter-band distance depends on the separation between 2B moieties. 
 
\section{Methods}

\begin{scheme}[b]
	\centering
	\includegraphics[width=1\columnwidth]{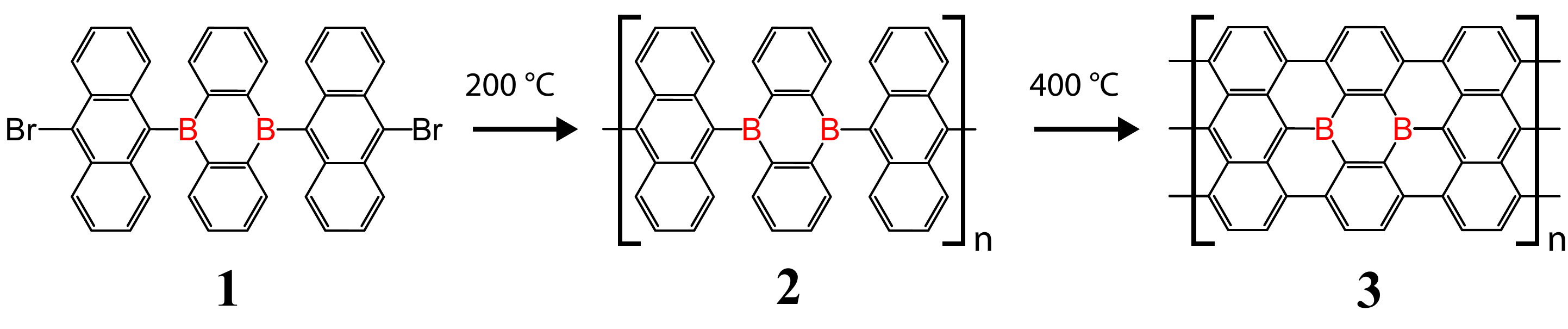}
	\caption{Chemical structure  of the precursor 9,10-bis(10-bromoanthracen-9-yl)-9,10-dihydro-9,10-diboraanthracene (\textbf{1}), and the two-step reaction pathway undergone on the Au(111) surface: the Ulmann-like C-C coupling into extended  polymers (\textbf{2}), and the cyclodehydrogenation into the flat  di-boron-doped graphene nanoribbon 2B-7-AGNRs  (\textbf{3}). }
	\label{precursor}
\end{scheme}

\textbf{SPM methods}. Our STS measurements were performed in an ultra-high vacuum system, composed of a chamber for the sample preparation and an independent chamber hosting a custom-made low temperature (4.8 K) Scanning Tunneling Microscope (STM). We measured the differential conductance using a lock-in demodulation method. 

\textbf{Angle-resolved photoemission measurements}. ARPES measurements were performed in a separate UHV system equipped with a room temperature STM and a low-energy electron diffraction (LEED) system used to check the sample quality.  ARPES data were taken using a Phoibos 150 SPECS high-resolution hemispherical electron analyzer on the sample cooled down to 150 K. He-I (h$\nu$ = 21.2 eV) radiation was provided by a high intensity UVS-300 SPECS discharge lamp coupled to a TMM-302 SPECS monochromator. Binding energies values given here are referred to E$_{\rm{F}}$. Because ARPES is an ensemble-averaging technique, a measurement of the electronic dispersion along the ribbons requires uniaxially aligned GNRs. Therefore, we used  the stepped Au(322) substrate instead of the Au(111), which consists of narrow terraces with the Au(111) orientation with  parallel steps separated by $\sim$1.2 nm.  This surface is a template for the growth of GNRs uniaxially aligned  along the narrow terraces \cite{MerinoJPCL18}. The dense array of steps have the additional effect of lowering the work function of the Au(111) surface by 0.25 eV  \cite{MerinoJPCL18}.

\textbf{Sample preparation.} On both the Au(111) and the Au(322) surfaces, the boron-doped graphene nanoribbons 2B-7-AGNRs, were obtained by following a two-step on-surface reaction of \textbf{1} (Scheme 1) precursors, as previously reported in Refs.\  \onlinecite{Cai2010,cloke_site-specific_2015,kawai_atomically_2015}. First,  a clean Au(111) single crystal surface was precovered with precursor \textbf{1} and annealed to 200 $^\circ$C \np{for $\sim$ 5 minutes} to activate their Ullmann-like coupling and polymerization into structure \textbf{2}. Further annealing to  400 $^\circ$C \np{($\sim$3 minutes)} induced a cyclodehydrogenation reaction, leading to additional C-C bonds between neighbor anthracene units and the creation of the flat GNR backbone \textbf{3}. As shown in Scheme \textbf{1}, the two boron atoms of the precursor appear at the center of the ribbon, a pair every 3-unit cells of the pristine 7-AGNR ribbon, amounting to a total $\sim$5$\%$ of the total C atoms of the ribbon.

\textbf{Theoretical methods.} \textit{Ab initio} calculations were carried out using DFT, as implemented in the SIESTA code~\cite{Sol02}. For the description of the 2B-7-AGNR adsorbed on gold, we used a supercell composed of a slab containing 3 layers of the Au(111) surface and the  nanoribbon on  top. We employed the stacking geometry suggested in Ref.~\onlinecite{giovannetti_PRL2008}, where $2 \times 2$ graphene and Au(111) $\sqrt 3 \times \sqrt 3$ R30$^\circ$ unit cells are directly matched. Following Ref.~\onlinecite{gonzalez-lakunza_PRL2008}, the bottom Au surface was passivated with hydrogen atoms to quench  the Au(111)'s  surface state of the bottom surface, avoiding spurious effects due to the interaction of the surface states on both sides of the slab.  Moreover, to avoid interactions between periodic surfaces, we considered a vacuum region of more than 20 \AA. 

The GNR and the top two Au layers were fully relaxed until forces were $<0.01$ eV/\AA. Dispersion interactions were taken into account using the nonlocal optB88-vdW functional \cite{klimes_JPCM2010}. The basis set consisted of double-$\zeta$ plus polarization (DZP) orbitals for C, B, H, and bulk Au atoms. Au atoms on the topmost layer were treated with a DZP basis set optimized for the description of the (111) surface of Au \cite{garciagil_PRB2009}. This extended, optimized basis set describes adequately the decay of the metal electron density into vacuum, and improves the description of the Au(111) surface state. A $3 \times 1 \times  1$ Monkhorst-Pack mesh was used for the k-point sampling of the three-dimensional Brillouin zone, where the 3 k-points are taken along the direction of the ribbon. A cutoff of 300 Ry was used for the real-space grid integrations. 
For the free standing 2B-7-AGNR,
we performed DFT calculations (including geometrical optimization) using the same   simulation  parameters and the same unit cell as for the adsorbed ribbon. This allows us to directly compare the band structures obtained in the isolated and the Au-supported cases. \np{All the DFT results presented in this article are spin-restrictred simmulations, and are qualitatively identical to similar calculations including spin as an additional degree of freedom. So, even for the free ribbon, the system does not show any net spin-polarization according ot our DFT simulations. }

\section{Results and Discussion}

\begin{figure} [tb!]
	\centering
	\includegraphics[width=0.99\columnwidth]{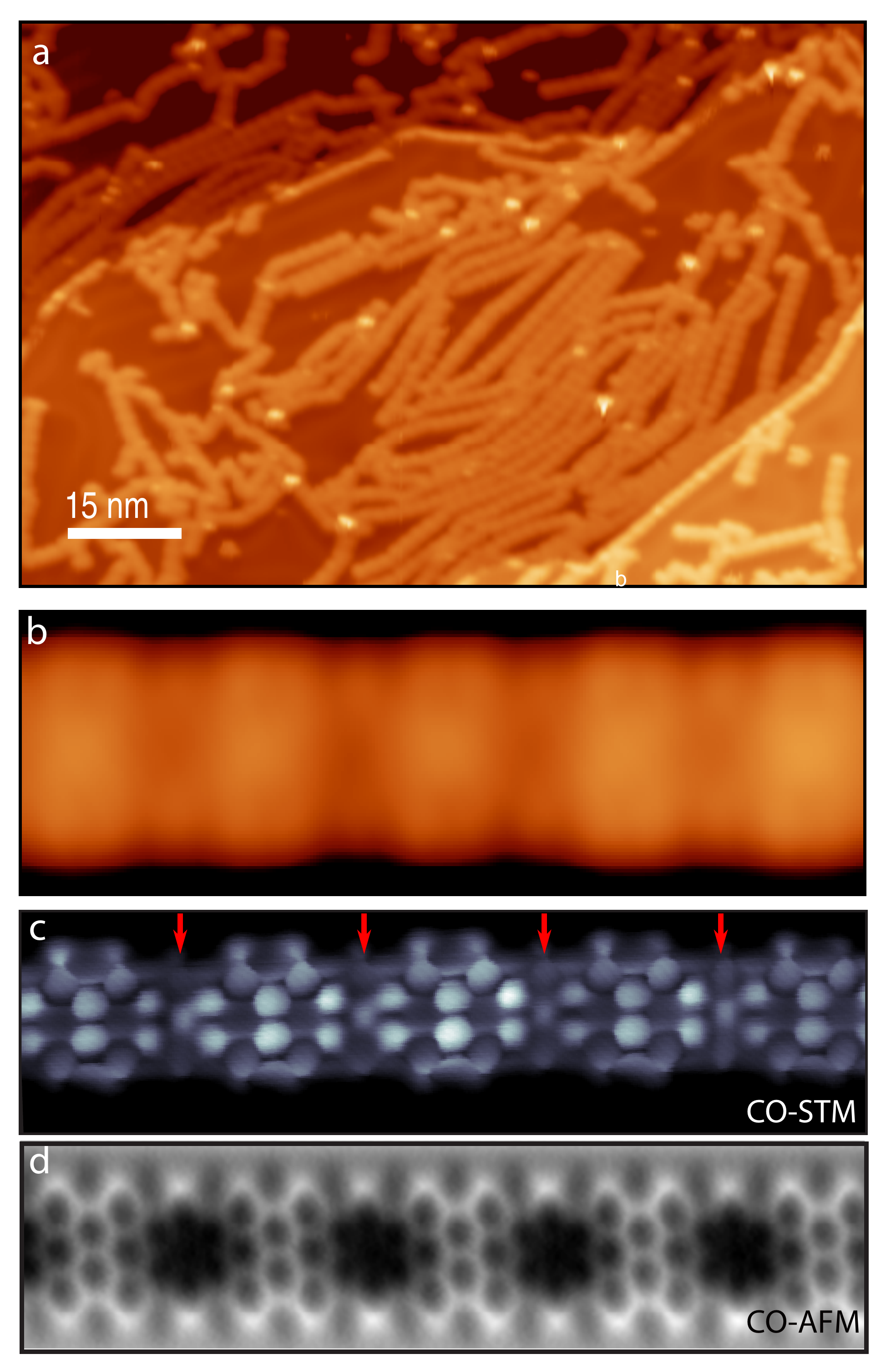}
	\caption{(a) Overview of the 2B-7-AGNR on Au(111) (\vs=1 V \It=170 pA). (b) Small scale STM image of a 2B-7-AGNR where the lower topography regions correspond to the borylated areas (\vs=-150 mV \It=410 pA). (c-d) High resolution dI/dV (c) and AFM (d) images taken with a CO functionalized tip in constant height mode. Red arrows highlight the borylated segments. %The scale bar in (c) is 10 \AA, and is also valid for (b) and (d). 
    }
	\label{Figure1}
\end{figure}

\begin{figure} [t!]
	\centering
	\includegraphics[width=0.99\columnwidth]{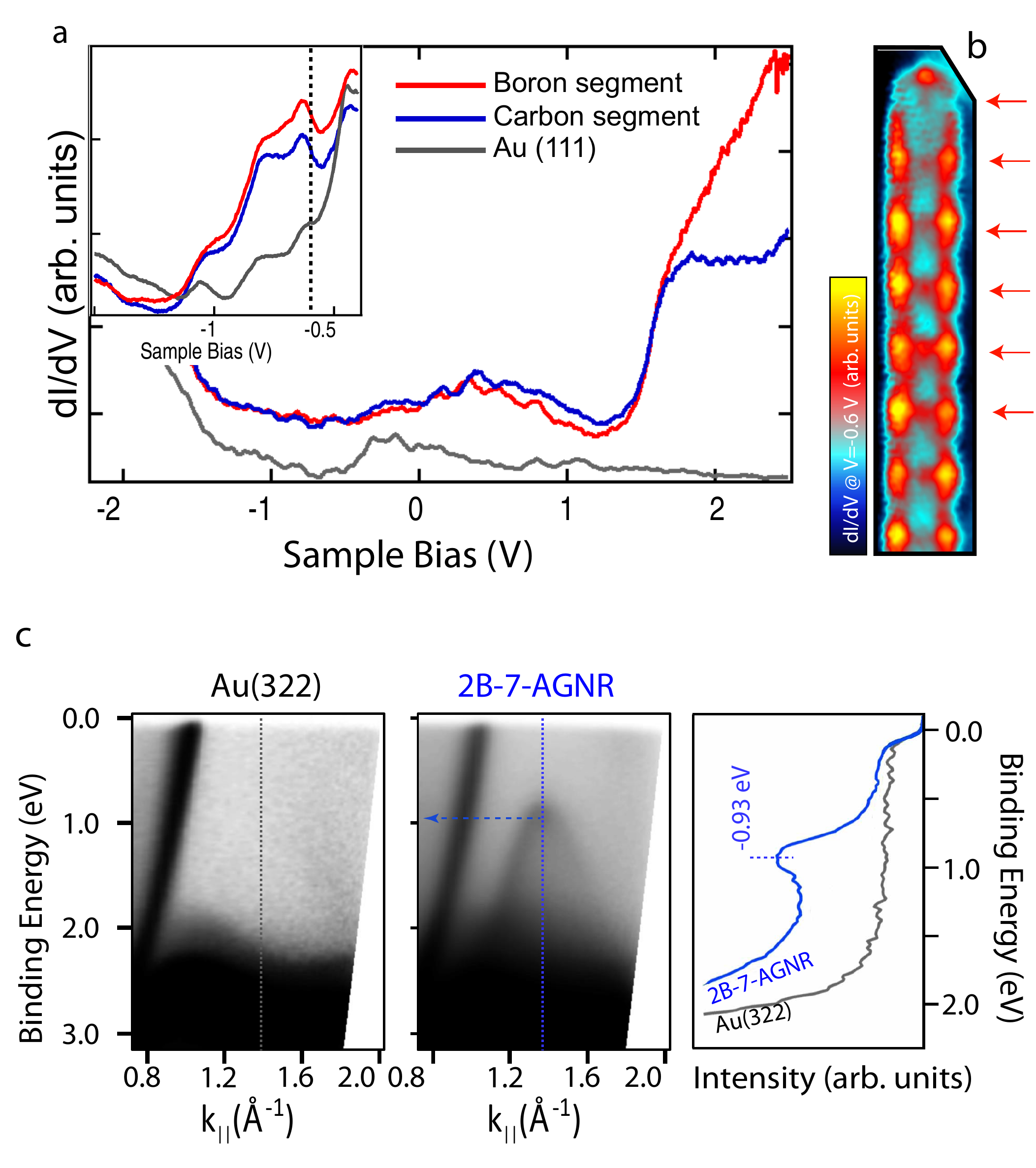}
	\caption{  (a) \didv\ spectra acquired at the edge of a borylated (red) and of a carbon (blue) segment (spectrum in gray is measured over the Au(111) as a reference). A dI/dV step  at $\sim$1.6 eV  is attributed to an \np{unoccupied} band. We associate a broad feature between 0.2 and 1 V  to the distorted surface state hybridized with an impurity  band (see text and Fig. 3). 
The inset shows a constant current (closed feedback) dI/dV spectra acquired at the edges of the ribbon, resolving a weak onset of conductance at $\sim$-0.6 V, \np{which is attributed to the valence band.} (Open feedback conditions \Vs=-1.1 V, \It=0.4 nA, V$_{\textrm{rms}}$=18 mV, \textit{f}=751.9 Hz; Inset: closed feedback conditions \It=0.5 nA,V$_{\textrm{rms}}$=12 mV, \textit{f}=769 Hz) (b) Constant current dI/dV map measured at the bias value of the dI/dV onset, at V=-0.6 V. Arrows mark the position of the borylated units. (c) Angle-resolved photoemission spectral maps of the bare Au(322) stepped surface and of 2B-7-AGNRs grown on a Au(322), where  the ribbons align parallel to the steps. On the right, ARPES spectra at the maximum of the 2B-7-AGNRs band and of the clean surface are plotted. }	\label{Figure2}  
\end{figure}

Figure \ref{Figure1}a shows an overview of the resulting ribbons  on the Au(111) surface. As reported in Refs.\ \onlinecite{kawai_atomically_2015,cloke_site-specific_2015}, the on-surface reaction of precursor \textbf{1} is complete, and produces extended ribbons with a small internal contrast in STM images.  A close-up view on a ribbon segment (Fig.\ \ref{Figure1}b) highlights the distinctive topographic contrast of 2B-7-AGNRs, where borylated regions appear as slight depressions along the ribbon backbone. To enhance the resolution of the ribbon we functionalized the STM tip  with a CO molecule \cite{gross_high-resolution_2011} and mapped the tunneling current in a constant height image at very low bias \cite{Hieulle2018}. This  method achieves high resolution images of the ring structure of the graphenoid backbone (Fig.\ 1c), which are comparable with constant-height AFM images (Fig.\ 1d). Interestingly, while the boron-regions are imaged as dark areas in AFM, they can be resolved in CO-STM, but with a lower contrast (red arrows in Fig.\ 1c). Although both imaging methods have some sensitivity to electronic states, they suggest that the borylated regions are lower \cite{kawai_atomically_2015} and, consequently, hint towards a stronger hybridization of boron atoms with the gold surface.
 
\textbf{Valence band determination}. We performed STS measurements on 2B-7-AGNR to determine the impact of the boron substitution on the 7-AGNR electronic structure. Figure \ref{Figure2}a shows the \didv\ spectra obtained at the edges of the borylated (red) and carbon (blue) regions of the 2B-7-AGNRs and the reference spectrum of the clean Au(111) surface (grey). The most prominent feature is the presence of a strong onset at $\sim$1.6 V, similar to the case of the conduction bands in pristine 7-AGNR \cite{ruffieux_electronic_2012,sode_electronic_2015}. Thus, we attribute this onset  to the conduction band of the borylated ribbon CB$^\text{B}$. \np{The identical alignment of this band than in the pristine case suggests that no significant charge transfer with the gold surface is to be expected in the borylated ribbon.}
Above this onset %(from now on labelled CB), 
the conductance in the carbon segments reaches a plateau, while on the borylated segments we detect a lager density of states at higher energies. 

In contrast to pristine 7-AGNRs, the onset of the valence band (VB) is not easily distinguishable in  \didv\ spectra of 2B-7-AGNRs \cite{kawai_atomically_2015,cloke_site-specific_2015}. However, we found evidence of the presence of occupied frontier bands  from dI/dV spectra in constant-current (closed feedback)  mode.  The spectra (inset of Fig.\ 2a) shows an onset of dI/dV signal at $\sim$-0.6 V, which extends to bias values below -1 V, both at the carbon and the borylated regions. The dI/dV signal associated to this band appears stronger at the edges and also shows some higher signal over the boron sites (Fig.\ 2b). However, this is far from representing the band's real shape  due to spurious effects regarding the different extension of the GNR orbitals into the vacuum \cite{sode_electronic_2015}. 
As we shall show later, the probable origin of the weak valence signal in dI/dV spectra as compared to the pristine version of 7-AGNR is a band order reversal with respect to the 7-AGNR band structure. This reversal is induced by boron states that selectively mix with the carbon $\pi$-system,  as we detailed in Ref.\ \onlinecite{carbonell-sanroma_quantum_2017}.

\begin{figure*} [th!]
	\centering
	\includegraphics[width=.70\textwidth]{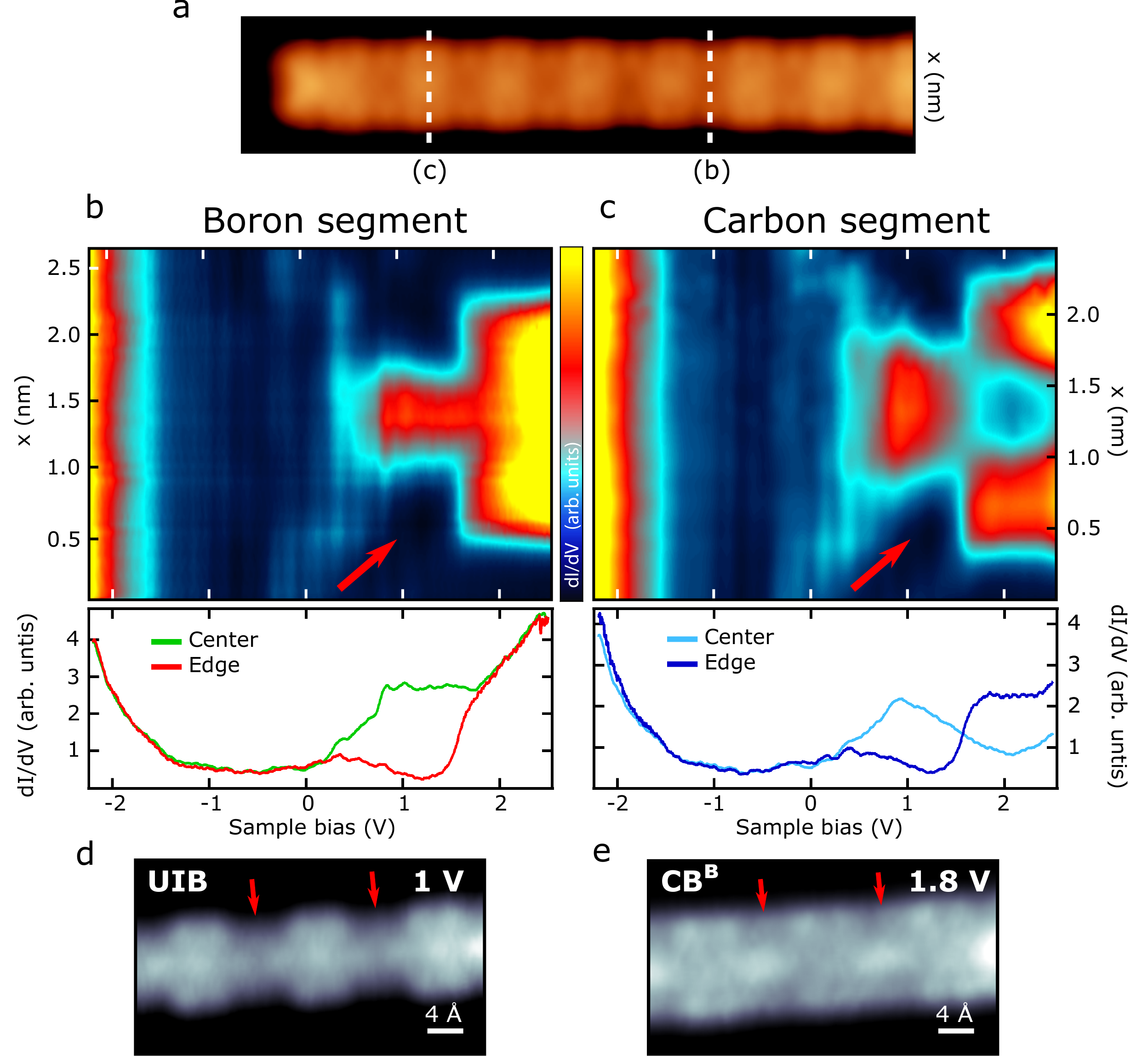}
	\caption{ (a) STM topography of a 2B-7-AGNR. (\vs=-150 mV \It=410 pA)  \didv\ spectral maps taken across (b) a boron segment and (c) a carbon segment of the  2B-7-AGNR (dashed lines in (a)), including dI/dV plots at the edge  and at the center  of the ribbon. The UIB is  localized at the center of the ribbon, while the CB$^\text{B}$ appears stronger at the edges.   The Au(111) SS is visible outside the ribbon at -500mV and  shifts upward underneath (following the red arrows). Constant height \didv\ maps of  (d) the UIB and (e) the CB$^\text{B}$, measured by approaching a CO-functionalized tip by 0.8 \AA\ and 1.2 \AA, respectively, from the open feedback conditions \Vs= 20 mV, \It= 100 pA.  Red arrows mark the position of the borylated regions (V$_{\textrm{rms}}$=14 mV, \textit{f}=760 Hz).}
	\label{2BSTSacross}
\end{figure*}

To explore the region of occupied bands of the borylated nanoribbons we also performed angle-resolved photoelectron spectroscopy of 2B-7-AGNRs on Au(322). The phoemission results resolve a clear dispersing band with an effective mass of 0.12 m$_\text{e}$ and its onset at a binding energy of  0.93 eV (Fig.\ 2c). We note that the work function of this surface is 0.25 eV lower than that of Au(111)  \cite{MerinoJPCL18}. Hence, if we consider that for weakly interacting adsorbates the bands shift according to changes in the work-function, these results  would correspond to a band's onset energy of  0.68 eV on the Au(111) surface. This value is close to the value observed in STS, although the low effective mass of the band´s electrons is intriguing. As we shall show later, DFT finds that this band corresponds instead to the VB-1  of the pristine ribbon, which is however not visible in  STS due to its nodal-plane structure \cite{sode_electronic_2015,talirz_-surface_2017,Senkovskiy2018}.

\textbf{Impurity states of 2B-AGNR.} The \didv\ spectra in Fig.\ 2a also show the presence of a broad conductance signal, between  E$_\text{F}$ and  CB$^\text{B}$, which is absent in the electronic structure of the pristine 7-AGNR. To explore in more detail the origin of this broad feature, we present in Fig.\ \ref{2BSTSacross}b and \ref{2BSTSacross}c two  stack plots of \didv\ spectra (spectral map) taken across the borylated  and the carbon sections of the 2B-7-AGNR, respectively (Figure \ref{2BSTSacross}a). The spectral maps highlight the presence of a broad feature above 0.3 V, with maximum intensity at around 1 V, i.e.\ within the CB$^\text{B}$-VB$^\text{B}$ band-gap, strongly localized at the center of the ribbon,  and extending along  the ribbon's axis.  Its energy and spatial distribution suggest that it corresponds to a  new unoccupied impurity band (UIB) originated by the boron substitution.  A similar feature was denominated acceptor band in Ref.\ \onlinecite{cloke_site-specific_2015}. 

Interestingly, the spectral maps of Fig.\ \ref{2BSTSacross} show an evolution of features attributed to the Au(111) surface state (SS) in the  proximity of the ribbon (red arrows in Figures \ref{2BSTSacross}b and c). The SS signatures gradually shift and merge with the UIB. This effect suggests  a sizable hybridization of boron-rich states with the SS, which is consistent with the broad line width of the UIB. 

Constant height \didv\ maps at the  UIB's maximum signal (1\,V) and the CB$^\text{B}$'s onset show distinct modulations, both of which  detail the absence of a nodal plane along  the AGNR axis, and thus evidences the even symmetry of these bands across the ribbon (Fig.\ \ref{2BSTSacross}d and \ref{2BSTSacross}e).  Furthermore, the  edges of the ribbon are depleted of states at the borylated segments in both bands, whereas the carbon segments accumulate states, specially for the UIB-derived band. In spite of the hybridization with the SS, the dI/dV maps agree fairly well with the simulated local density of states (LDOS) maps for the unoccupied bands of an   ribbon adsorbed on the Au(111) surface (shown in Fig.\ 4d-4f). 
%As we will describe later, our findings indicate that the  CB$^\text{B}$ corresponds to the (CB+1) band of the pristine ribbon.

\begin{figure*} [t!]
\centering
\includegraphics[width=.99\textwidth]{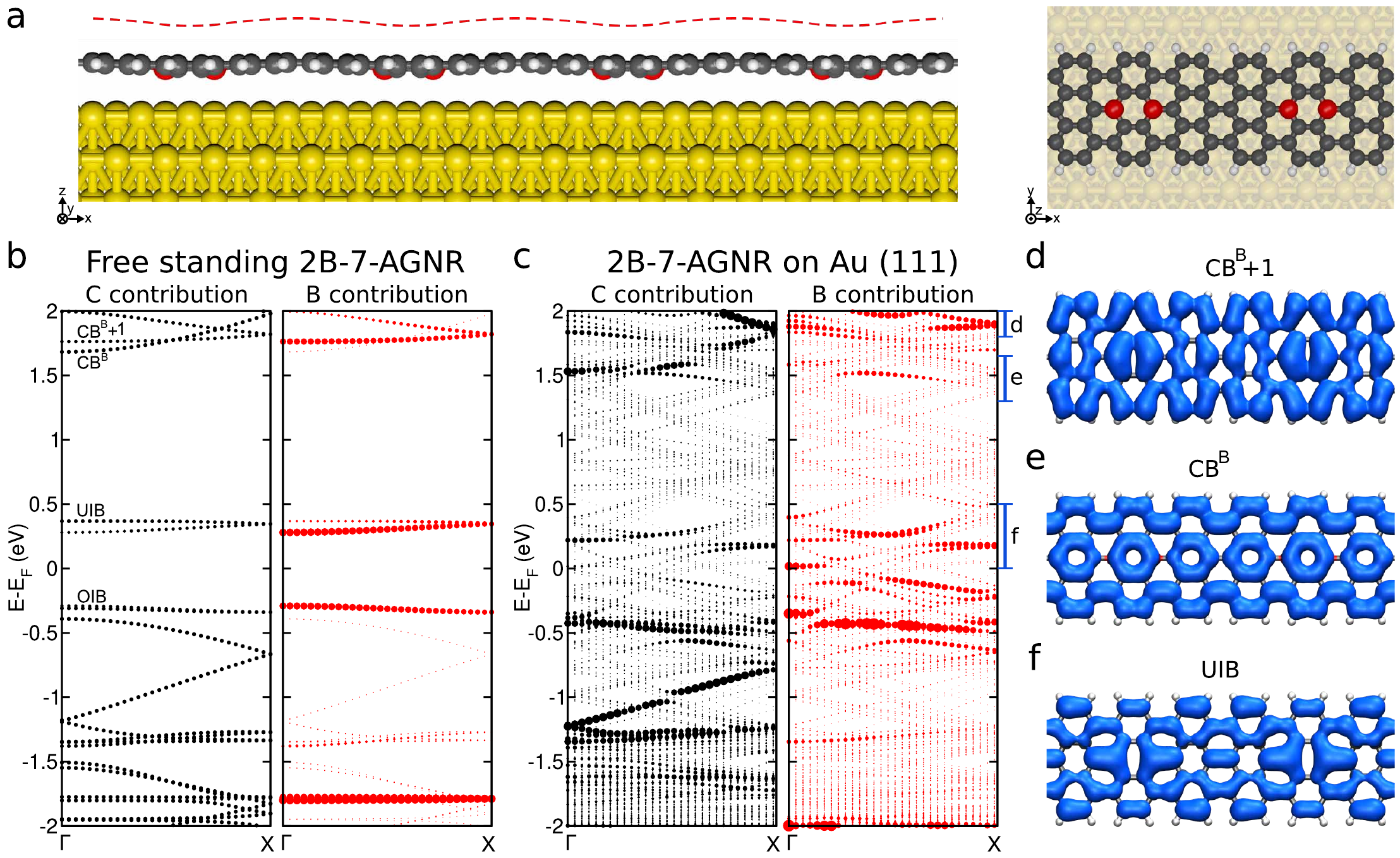}

\caption{
(a) Side view of a 2B-7-AGNR on a Au(111) slab after geometry optimization (dashed red line indicates the obtained profile using the atoms in the GNR backbone, where the borylated region are 0.3~\AA\ closer to the metallic surface). A top view of the relaxed structure is shown on the right. (b) Band structure of a free-standing 2B-7-AGNR \np{along the $\Gamma-$X direction  (X-direction as indicated in (a)). Left}  and right panels present
the carbon and boron character of the bands, respectively. 
The radius of the black (red) circles represent the C (B) contribution of each electronic state.
Two clear boron heavy bands appear at $\pm$0.3 eV with respect to E$_{\rm{F}}$, which for the free standing nanoribbon is arbitrarily defined to be in the middle of the gap between them. For visualization purposes, the C and B contributions have been multiplied by 1/3 and 3, respectively. (c) Same as in \textit{b} but for 2B-7-AGNR on top of the gold slab, with the B contributions multiplied by 10.  (d-f) \np{Constant LDOS isosurfaces (0.0004 e$^-$/Bohr$^3$)} of the CB$^\text{B}$, CB$^\text{B}$+1, and UIB bands for the Au-supported 2B-7-AGNR. The considered energy range of integration for each isosurface is indicated on the right side of panel c. In the caption of Fig.~\ref{bands_wfs}  refer the labeling scheme of the bands to those of the pristine ribbon.
} 
\label{2BBands}
\end{figure*}

\textbf{Density Functional Theory results.} A better understanding of the electronic structure of the boron doped nanoribbons and their interaction with the underlying metal substrate is provided by \emph{ab initio} simulations. 
In particular, we performed DFT calculations for 2B-7-AGNRs, both free standing and adsorbed on a Au(111) substrate. 
In the latter case, special care was taken to ensure a correct description of the Shockley  state of the metal surface (more details in the methods section above). The optimized geometry of the 2B-7-AGNR/Au(111) system is shown in Figure \ref{2BBands}a.  The presence of the boron pairs induces a significant deformation of the nanoribbon: the borylated regions appear $\sim$0.3 \AA\ closer to the metal surface, in close agreement with the results from Ref.\ \onlinecite{kawai_atomically_2015}. This deformation is absent in the free ribbon, and hence, reflects a sizable nanoribbon-metal hybridization.
In the following we describe the main ingredients that characterize the band structure of the 2B-7-AGNRs.

{\it In-gap impurity bands:} Figure~\ref{2BBands}b shows the band structure of the free-standing nanoribbon projected onto carbon and boron orbitals, thus reflecting the contribution from different species to each band. Notice that this band structure has been computed using a unit cell that contains two 2B-defects in order to facilitate the comparison with the adsorbed case (Fig.~\ref{2BBands}a).  
The most characteristic feature is a pair of bands with large boron contribution, located close to the Fermi level ($\sim\pm$0.3~eV in Fig.~\ref{2BBands}b). They correspond to an occupied impurity-induced band (OIB) and an unoccupied impurity-induced band (UIB). The presence of these two bands at E$_{\rm{F}}$  was already identified at low doping concentration in Ref.~\onlinecite{carbonell-sanroma_quantum_2017} (also shown in Fig. 5). Intriguingly, very little B content is present in the bands immediately below and above these impurity bands, i.e. in the delocalized frontier bands of the 2B-7-AGNR, whose origin will be discussed later in more detail. We also find an additional band with large  B contribution at $\sim$~-1.9~eV. As in the case of both in-gap impurity states, this band is rather flat, denoting its large degree of localization at the 2B sites.  This band was also identified in the high dilution limit in Ref.~\onlinecite{carbonell-sanroma_quantum_2017}

\begin{figure*}[t!]
   \centering
   \includegraphics[width=.90\textwidth]{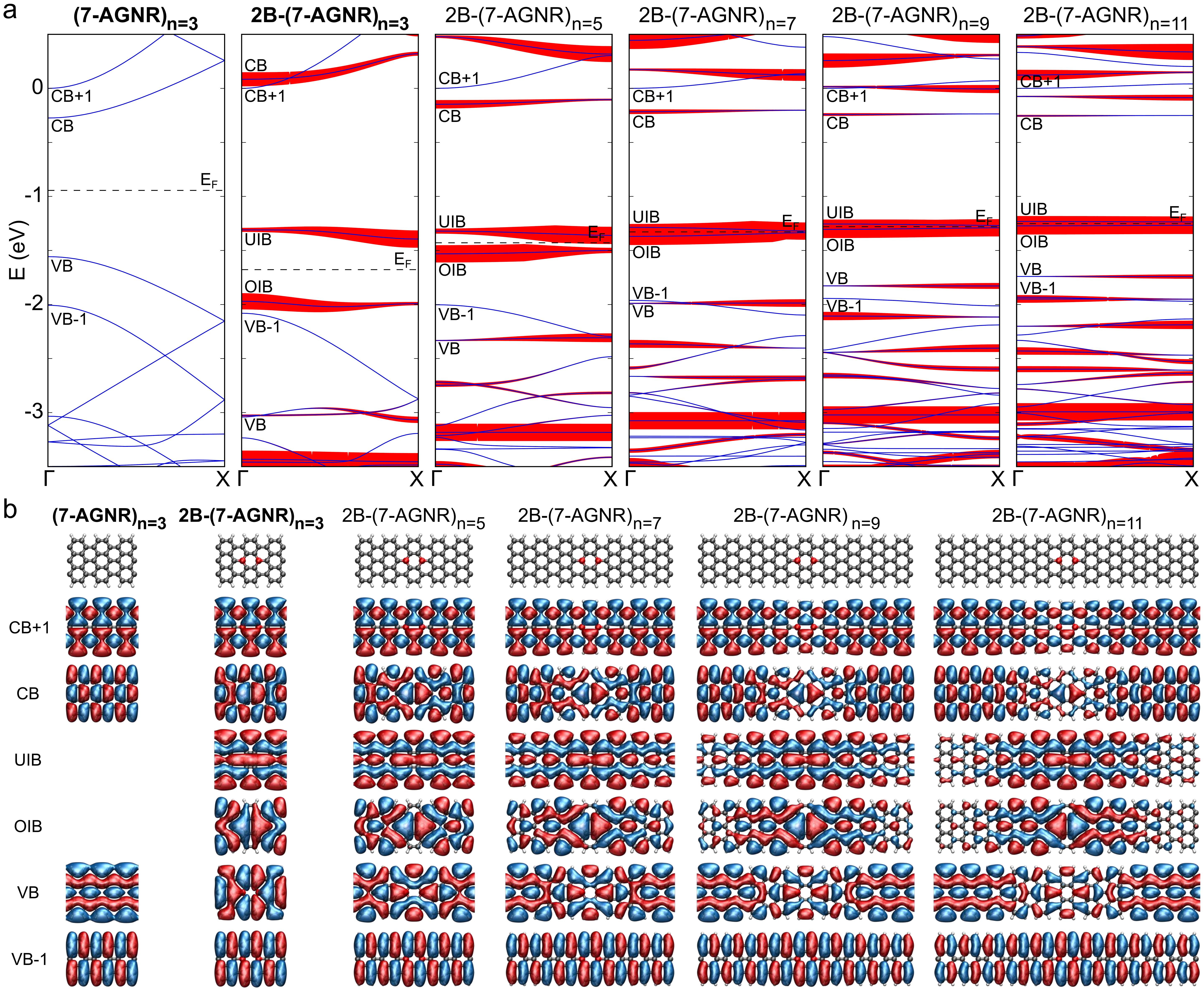}
   \caption{ (a) Band structure of a free-standing 7-AGNR (left) and of 2B-7-AGNRs with decreasing doping concentrations (from left-to-right) reflected by the increase of the unit cell size containing a single 2B impurity.  The size of the red shaded area around a given band is proportional to its boron character. The labels of the 7-AGNR frontier bands VB-1/VB/CB/CB+1, are assigned to the borylated ribbons according to their wave-function's amplitude isosurfaces, shown in (b), such that CB$^\text{B}$=CB+1, CB$^\text{B}$+1=CB, VB$^\text{B}$=VB-1, and VB$^\text{B}$-1=VB, thus defining the reversal of frontier bands explained in the text.  The  occupied/unoccupied mid-gap impurity-induced bands are labeled as OIB/UIB. The Fermi energy is arbitrarily assigned in the middle of the  gap between the highest occupied and lowest unoccupied bands,  and all the band structures are aligned with respect to CB+1. (b) Representation of the unit cell (top) and the real part of the wave-functions at $\Gamma$ corresponding to the frontier and impurity-induced bands marked in (a). \np{The color indicate the phase of the wave function, whereas the surfaces are constant amplitude (0.01) isosurfaces.} }
   \label{bands_wfs}
\end{figure*}

These two impurity bands can be rationalized by considering that boron states of neighboring 2B sites hybridize with each other and form these flat bands. For very low impurity concentration in a free-standing ribbon (e.g. large separation between 2B dopants, see Supporting Information in Ref.~\onlinecite{carbonell-sanroma_quantum_2017}), two boron states lie very close to E$_{\rm{F}}$  with a shape that resembles the impurity bands of the present case. In Fig.\ \ref{bands_wfs}a we show the evolution of the band structure as the borylated  moieties are approached. 
The two boron states found at  E$_{\rm{F}}$ for the diluted case (n=11) gradually   increase their energy spacing as the 
boron moieties become closer than 7 unit cells  of the pristine 7-AGNR, and  develop a weak dispersion.  
Figure 5b shows the evolution of the wave functions of the localized boron states. As the 2B sites get  closer the impurities' wave functions  overlap and become the new impurity bands (OIB and UIB) of the current (n=3) 2B-7-AGNR. From these results, we foresee that the OIB-UIB energy spacing can be tuned by adjusting the length of the carbon segments used for the on-surface reaction.

Interestingly, the wave-functions for the OIB and UIB of the 2B-7-AGNR (obtained at the $\Gamma$ point, shown in the second column of Figure~\ref{bands_wfs} b)  exhibit an even symmetry across the ribbon for both impurity states~\cite{carbonell-sanroma_quantum_2017}, in agreement with dI/dV maps of Fig.~\ref{2BSTSacross}. At the $\Gamma$ point (Fig.~\ref{bands_wfs}) the OIB has a nodal plane in the middle of the 2B-defect, while UIB presents even parity with respect to the 2B-defect center. This contrasts with the apparent nodal plane observed in the experimental map of the UIB (Fig.~\ref{2BSTSacross}d). The situation, however, is reversed as we move towards the Brillouin zone boundary (not shown in the figure). Hence,  a direct comparison of experimental STS maps with wave-function isosurfaces of periodic systems at a particular $\bf{k}$ point can lead to errors in the band assignment. We thus  plot in Fig.~\ref{2BBands}(d-f) the computed \agln{LDOS} maps of the unoccupied bands for the 2B-7-AGNR on the Au(111) surface, which can be more directly correlated with the dI/dV maps of Fig.~\ref{2BSTSacross}. The even symmetry across the ribbon is maintained  for the B-rich bands, while the low signal between the two B atoms of the UIB is now clearly reproduced by the simulations.

{\it Band-reversal at the band-gap edges:} Surprisingly, the original VB of the pristine 7-AGNR is absent in the band structure of the 2B-7-AGNR in Fig.~\ref{2BBands}b. Instead, the first carbon-rich ribbon state is a dispersing band, slightly below the OIB, which has the characteristic wave-function (plotted at $\Gamma$ in Fig.~\ref{bands_wfs}b) of the VB-1 of the pristine ribbon.
This band is not affected by the presence of the boron impurities because its has odd symmetry across the ribbon axis~\cite{carbonell-sanroma_quantum_2017}, and presents a clear nodal plane along the axis, where the 2B moieties lie.
The negligible interaction of VB-1 with the boron impurities is 
evident from Fig.\ \ref{bands_wfs}b, where the corresponding wave-function is identical to that of the pristine ribbon and not affected by the change in the separation between the boron moieties. 
Additionally, this band has many nodal-planes  along the ribbon, which explains why it is difficult to detect in normal STS measurements, as it is the case for the VB-1 of the pristine ribbons \cite{sode_electronic_2015}.  

In contrast, the original VB  is known to interact strongly with boron states because both have  even symmetry across the ribbon.  As previously reported\cite{carbonell-sanroma_quantum_2017}, this band is strongly scattered by 2B moieties and confined in the carbon 7-AGNR segments between them, appearing as quantized flat bands at lower energies. The quantized VB of the pristine ribbon appears in Fig.\ \ref{bands_wfs}b, for the n=11 case, split in several modes with large B content. The VB band onset  shifts down significantly as the distance between 2B moieties (i.e. the confined region) becomes smaller. For the n=3 case, the first VB derived QWS appears in 
Fig.~\ref{2BBands}b as a quite flat band at $\sim$~-1.3~eV with a moderate B content that indicates its hybridization with the 2B impurities.

The results in Fig.\ \ref{bands_wfs} also confirm that a
similar reversal of frontier bands due to symmetry-selective band mixing applies to the conduction bands. The original CB of 7-AGNR is even across the ribbon and, thus, interacts with the 2B states, in contrast to the pristine CB+1, which is odd.  However, the interaction is weaker in this case,  and the pristine CB moves to lie slightly above the CB+1, which becomes the new frontier band. In contrast to the case of valence bands, the pristine CB was in this case the ``hidden" band in STS spectra because of the numerous nodal planes along the ribbon \cite{sode_electronic_2015}, and the one usually detected was the CB+1, which has odd symmetry across and is not affected by the boron. This explain that in both pristine and borylated ribbons, the conduction band detected by STM, the CB+1, lies at the same position.

{\it Adsorption on the Au(111) surface:} When adsorbed on the Au(111) slab, the main ingredients of the electronic structure of the free-standing 2B-7-AGNRs can be still identified. One of the most significant changes though, is the mixing of the impurity bands with Au states.
Figure \ref{2BBands}c shows that UIB   undergoes a larger distortion and mixture with Au states  upon adsorption, resulting in a broad band of boron-rich states that accounts for its spectral line shape   in our experiment. 

In contrast, the dispersion of the carbon-rich bands and the OIB can still be recognized in the band plots, signaling their weaker hybridization with the states at the surface. We observe a small mixing between the originally orthogonal  VB-1 and OIB, probably due to their proximity and the mediation of the metal. Since the OIB is a rather flat band, we conclude that the band resolved by ARPES in Fig.\ref{Figure2}c corresponds to the VB-1, while the one measured by STM in Fig.~\ref{Figure2}a is probably the OIB, because it appears modulated with the periodicity of the boron subunits.

\section{Conclusions}

In conclusion, we have studied the structural and electronic effects of a boron-substitution on armchair graphene nanoribbons. The resulting structures exhibit a high boron-dopant concentration and a well-defined periodicity. The most prominent change on the electronic structure of 2B-7-AGNRs is the introduction of two in-gap impurity
bands, which arise from the overlap of nearby borylated moieties, and mix strongly with Au(111) surface states resulting in 
hybrid metal-ribbon states.  The 2B dopants also cause a reversal of the carbon-rich bands at the gap edges, compared to the pristine 7-AGNR ribbon. This is caused by the symmetry selective mixing of bands with the states of the di-boron defect, which leads to quantum confinement of original VB and CB and their shift away from the gap. The new valence and conductance bands have a different symmetry both across and along the ribbon. Indeed, they correspond to the VB-1 and CB+1 of the pristine 7-AGNR. This is the reason for the previously reported \cite{kawai_atomically_2015,cloke_site-specific_2015} disappearance of the VB from STS spectra. 
Our results further define basic rules for tuning the band structure of the ribbons: varying  the distance between 2B groups can be employed to tune the gap between impurity bands; rotation of the 2B moiety with respect to the ribbon axis changes the symmetry of the carbon bands that they mix with, thus affecting the character of the frontier transport bands.

%%%%%%%%%%%%%%%%%%%%%%%%%%%%%%%%%%%%%%%%%%%%%%%%%%%%%%%%%%%%%%%%%%%%%
\begin{acknowledgement}
We acknowledge financial support from the  Basque Government (Departamento de  Industria, grant no.
PI-2015-1-42,  and Departamento de Educaci\'on and UPV/EHU, grant No. IT-756-13), and
the Spanish Ministerio de Econom\'{\i}a y Competitividad (Grants No. MAT2016-78293-C6, FIS2015-62538-ERC, FIS2017-83780-P), and the Maria de Maeztu Units of Excellence Programme MDM-2016-0618), the European Research Council
(grant agreement no. 635919), and the European Regional Development Fund.
\end{acknowledgement}

\bibliography{Bibliography_new}

\providecommand{\latin}[1]{#1}
\makeatletter
\providecommand{\doi}
  {\begingroup\let\do\@makeother\dospecials
  \catcode`\{=1 \catcode`\}=2\doi@aux}
\providecommand{\doi@aux}[1]{\endgroup\texttt{#1}}
\makeatother
\providecommand*\mcitethebibliography{\thebibliography}
\csname @ifundefined\endcsname{endmcitethebibliography}
  {\let\endmcitethebibliography\endthebibliography}{}
\begin{mcitethebibliography}{36}
\providecommand*\natexlab[1]{#1}
\providecommand*\mciteSetBstSublistMode[1]{}
\providecommand*\mciteSetBstMaxWidthForm[2]{}
\providecommand*\mciteBstWouldAddEndPuncttrue
  {\def\EndOfBibitem{\unskip.}}
\providecommand*\mciteBstWouldAddEndPunctfalse
  {\let\EndOfBibitem\relax}
\providecommand*\mciteSetBstMidEndSepPunct[3]{}
\providecommand*\mciteSetBstSublistLabelBeginEnd[3]{}
\providecommand*\EndOfBibitem{}
\mciteSetBstSublistMode{f}
\mciteSetBstMaxWidthForm{subitem}{(\alph{mcitesubitemcount})}
\mciteSetBstSublistLabelBeginEnd
  {\mcitemaxwidthsubitemform\space}
  {\relax}
  {\relax}

\bibitem[Celis \latin{et~al.}(2016)Celis, Nair, Taleb-Ibrahimi, Conrad, Berger,
  {De Heer}, and Tejeda]{Celis2016}
Celis,~A.; Nair,~M.~N.; Taleb-Ibrahimi,~A.; Conrad,~E.~H.; Berger,~C.; {De
  Heer},~W.~A.; Tejeda,~A. {Graphene nanoribbons: Fabrication, properties and
  devices}. \emph{J. Phys. D. Appl. Phys.} \textbf{2016}, \emph{49},
  143001\relax
\mciteBstWouldAddEndPuncttrue
\mciteSetBstMidEndSepPunct{\mcitedefaultmidpunct}
{\mcitedefaultendpunct}{\mcitedefaultseppunct}\relax
\EndOfBibitem
\bibitem[Yang \latin{et~al.}(2007)Yang, Park, Son, Cohen, and Louie]{Yang2007}
Yang,~L.; Park,~C.-H.~H.; Son,~Y.-W.~W.; Cohen,~M.~L.; Louie,~S.~G.
  {Quasiparticle energies and band gaps in graphene nanoribbons}. \emph{Phys.
  Rev. Lett.} \textbf{2007}, \emph{99}, 6--9\relax
\mciteBstWouldAddEndPuncttrue
\mciteSetBstMidEndSepPunct{\mcitedefaultmidpunct}
{\mcitedefaultendpunct}{\mcitedefaultseppunct}\relax
\EndOfBibitem
\bibitem[Wakabayashi \latin{et~al.}(2010)Wakabayashi, Sasaki, Nakanishi, and
  Enoki]{Wakabayashi2010}
Wakabayashi,~K.; Sasaki,~K.-i.; Nakanishi,~T.; Enoki,~T. {Electronic states of
  graphene nanoribbons and analytical solutions}. \emph{Sci. Technol. Adv.
  Mater.} \textbf{2010}, \emph{11}, 054504\relax
\mciteBstWouldAddEndPuncttrue
\mciteSetBstMidEndSepPunct{\mcitedefaultmidpunct}
{\mcitedefaultendpunct}{\mcitedefaultseppunct}\relax
\EndOfBibitem
\bibitem[Cai \latin{et~al.}(2010)Cai, Ruffieux, Jaafar, Bieri, Braun,
  Blankenburg, Muoth, Seitsonen, Saleh, Feng, Mullen, and Fasel]{Cai2010}
Cai,~J.; Ruffieux,~P.; Jaafar,~R.; Bieri,~M.; Braun,~T.; Blankenburg,~S.;
  Muoth,~M.; Seitsonen,~A.~P.; Saleh,~M.; Feng,~X.; Mullen,~K.; Fasel,~R.
  {Atomically precise bottom-up fabrication of graphene nanoribbons}.
  \emph{Nature} \textbf{2010}, \emph{466}, 470--473\relax
\mciteBstWouldAddEndPuncttrue
\mciteSetBstMidEndSepPunct{\mcitedefaultmidpunct}
{\mcitedefaultendpunct}{\mcitedefaultseppunct}\relax
\EndOfBibitem
\bibitem[Talirz \latin{et~al.}(2016)Talirz, Ruffieux, and Fasel]{Tarliz2016rev}
Talirz,~L.; Ruffieux,~P.; Fasel,~R. {On-Surface Synthesis of Atomically Precise
  Graphene Nanoribbons}. \emph{Adv. Mater.} \textbf{2016}, 6222--6231\relax
\mciteBstWouldAddEndPuncttrue
\mciteSetBstMidEndSepPunct{\mcitedefaultmidpunct}
{\mcitedefaultendpunct}{\mcitedefaultseppunct}\relax
\EndOfBibitem
\bibitem[Corso \latin{et~al.}(2018)Corso, Carbonell-Sanrom\`a, and
  de~Oteyza]{corso_rev2018}
Corso,~M.; Carbonell-Sanrom\`a,~E.; de~Oteyza,~D. In \emph{On-surface synthesis
  II}; de~Oteyza,~D., Rogero,~C., Eds.; Springer: Switzerland, 2018; Chapter
  Bottom-up Fabrication of Atomically Precise Graphene Nanoribbons, pp
  113--152\relax
\mciteBstWouldAddEndPuncttrue
\mciteSetBstMidEndSepPunct{\mcitedefaultmidpunct}
{\mcitedefaultendpunct}{\mcitedefaultseppunct}\relax
\EndOfBibitem
\bibitem[Zhang \latin{et~al.}(2015)Zhang, Lin, Sun, Chen, Zagranyarski,
  Aghdassi, Duhm, Li, Zhong, Li, M\"ullen, Fuchs, and Chi]{zhang_-surface_2015}
Zhang,~H.; Lin,~H.; Sun,~K.; Chen,~L.; Zagranyarski,~Y.; Aghdassi,~N.;
  Duhm,~S.; Li,~Q.; Zhong,~D.; Li,~Y.; M\"ullen,~K.; Fuchs,~H.; Chi,~L.
  On-{Surface} {Synthesis} of {Rylene}-{Type} {Graphene} {Nanoribbons}.
  \emph{J. Am. Chem. Soc.} \textbf{2015}, \emph{137}, 4022--4025\relax
\mciteBstWouldAddEndPuncttrue
\mciteSetBstMidEndSepPunct{\mcitedefaultmidpunct}
{\mcitedefaultendpunct}{\mcitedefaultseppunct}\relax
\EndOfBibitem
\bibitem[Chen \latin{et~al.}(2013)Chen, de~Oteyza, Pedramrazi, Chen, Fischer,
  and Crommie]{chen_ACSNano2013}
Chen,~Y.-C.; de~Oteyza,~D.~G.; Pedramrazi,~Z.; Chen,~C.; Fischer,~F.~R.;
  Crommie,~M.~F. Tuning the Band Gap of Graphene Nanoribbons Synthesized from
  Molecular Precursors. \emph{ACS Nano} \textbf{2013}, \emph{7},
  6123--6128\relax
\mciteBstWouldAddEndPuncttrue
\mciteSetBstMidEndSepPunct{\mcitedefaultmidpunct}
{\mcitedefaultendpunct}{\mcitedefaultseppunct}\relax
\EndOfBibitem
\bibitem[Talirz \latin{et~al.}(2017)Talirz, S\"ode, Dumslaff, Wang,
  Sanchez-Valencia, Liu, Shinde, Pignedoli, Liang, Meunier, Plumb, Shi, Feng,
  Narita, M\"ullen, Fasel, and Ruffieux]{talirz_-surface_2017}
Talirz,~L. \latin{et~al.}  On-{Surface} {Synthesis} and {Characterization} of
  9-{Atom} {Wide} {Armchair} {Graphene} {Nanoribbons}. \emph{ACS Nano}
  \textbf{2017}, 1380--1388\relax
\mciteBstWouldAddEndPuncttrue
\mciteSetBstMidEndSepPunct{\mcitedefaultmidpunct}
{\mcitedefaultendpunct}{\mcitedefaultseppunct}\relax
\EndOfBibitem
\bibitem[Abdurakhmanova \latin{et~al.}(2014)Abdurakhmanova, Amsharov, Stepanow,
  Jansen, Kern, and Amsharov]{abdurakhmanova_Carbon2014}
Abdurakhmanova,~N.; Amsharov,~N.; Stepanow,~S.; Jansen,~M.; Kern,~K.;
  Amsharov,~K. Synthesis of wide atomically precise graphene nanoribbons from
  para-oligophenylene based molecular precursor. \emph{Carbon} \textbf{2014},
  \emph{77}, 1187 -- 1190\relax
\mciteBstWouldAddEndPuncttrue
\mciteSetBstMidEndSepPunct{\mcitedefaultmidpunct}
{\mcitedefaultendpunct}{\mcitedefaultseppunct}\relax
\EndOfBibitem
\bibitem[Merino-D{\'{i}}ez \latin{et~al.}(2017)Merino-D{\'{i}}ez, Garcia-Lekue,
  Carbonell-Sanrom{\`{a}}, Li, Corso, Colazzo, Sedona, S{\'{a}}nchez-Portal,
  Pascual, and {De Oteyza}]{MerinoACS17}
Merino-D{\'{i}}ez,~N.; Garcia-Lekue,~A.; Carbonell-Sanrom{\`{a}},~E.; Li,~J.;
  Corso,~M.; Colazzo,~L.; Sedona,~F.; S{\'{a}}nchez-Portal,~D.; Pascual,~J.~I.;
  {De Oteyza},~D.~G. {Width-Dependent Band Gap in Armchair Graphene Nanoribbons
  Reveals Fermi Level Pinning on Au(111)}. \emph{ACS Nano} \textbf{2017},
  \emph{11}, 11661--11668\relax
\mciteBstWouldAddEndPuncttrue
\mciteSetBstMidEndSepPunct{\mcitedefaultmidpunct}
{\mcitedefaultendpunct}{\mcitedefaultseppunct}\relax
\EndOfBibitem
\bibitem[de~Oteyza \latin{et~al.}(2016)de~Oteyza, Garc\'ia-Lekue, Vilas-Varela,
  Merino-D\'iez, Carbonell-Sanrom\`a, Corso, Vasseur, Rogero, Guiti\'an,
  Pascual, Ortega, Wakayama, and Pe\~na]{de_oteyza_substrate-independent_2016}
de~Oteyza,~D.~G.; Garc\'ia-Lekue,~A.; Vilas-Varela,~M.; Merino-D\'iez,~N.;
  Carbonell-Sanrom\`a,~E.; Corso,~M.; Vasseur,~G.; Rogero,~C.; Guiti\'an,~E.;
  Pascual,~J.~I.; Ortega,~J.~E.; Wakayama,~Y.; Pe\~na,~D.
  Substrate-{Independent} {Growth} of {Atomically} {Precise} {Chiral}
  {Graphene} {Nanoribbons}. \emph{ACS Nano} \textbf{2016}, \emph{10},
  9000--9008\relax
\mciteBstWouldAddEndPuncttrue
\mciteSetBstMidEndSepPunct{\mcitedefaultmidpunct}
{\mcitedefaultendpunct}{\mcitedefaultseppunct}\relax
\EndOfBibitem
\bibitem[Ruffieux \latin{et~al.}(2016)Ruffieux, Wang, Yang,
  S\'anchez-S\'anchez, Liu, Dienel, Talirz, Shinde, Pignedoli, Passerone,
  Dumslaff, Feng, M\"ullen, and Fasel]{ruffieux_-surface_2016}
Ruffieux,~P.; Wang,~S.; Yang,~B.; S\'anchez-S\'anchez,~C.; Liu,~J.; Dienel,~T.;
  Talirz,~L.; Shinde,~P.; Pignedoli,~C.~A.; Passerone,~D.; Dumslaff,~T.;
  Feng,~X.; M\"ullen,~K.; Fasel,~R. On-surface synthesis of graphene
  nanoribbons with zigzag edge topology. \emph{Nature} \textbf{2016},
  \emph{531}, 489--492\relax
\mciteBstWouldAddEndPuncttrue
\mciteSetBstMidEndSepPunct{\mcitedefaultmidpunct}
{\mcitedefaultendpunct}{\mcitedefaultseppunct}\relax
\EndOfBibitem
\bibitem[Nguyen \latin{et~al.}(2016)Nguyen, Toma, Cao, Pedramrazi, Chen, Rizzo,
  Joshi, Bronner, Chen, Favaro, Louie, Fischer, and
  Crommie]{nguyen_bottom-up_2016}
Nguyen,~G.~D.; Toma,~F.~M.; Cao,~T.; Pedramrazi,~Z.; Chen,~C.; Rizzo,~D.~J.;
  Joshi,~T.; Bronner,~C.; Chen,~Y.-C.; Favaro,~M.; Louie,~S.~G.;
  Fischer,~F.~R.; Crommie,~M.~F. Bottom-{Up} {Synthesis} of \textit{{N}} = 13
  {Sulfur}-{Doped} {Graphene} {Nanoribbons}. \emph{J. Phys. Chem. C}
  \textbf{2016}, \emph{120}, 2684--2687\relax
\mciteBstWouldAddEndPuncttrue
\mciteSetBstMidEndSepPunct{\mcitedefaultmidpunct}
{\mcitedefaultendpunct}{\mcitedefaultseppunct}\relax
\EndOfBibitem
\bibitem[Kawai \latin{et~al.}(2015)Kawai, Saito, Osumi, Yamaguchi, Foster,
  Spijker, and Meyer]{kawai_atomically_2015}
Kawai,~S.; Saito,~S.; Osumi,~S.; Yamaguchi,~S.; Foster,~A.~S.; Spijker,~P.;
  Meyer,~E. Atomically controlled substitutional boron-doping of graphene
  nanoribbons. \emph{Nat. Commun.} \textbf{2015}, \emph{6}, 8098\relax
\mciteBstWouldAddEndPuncttrue
\mciteSetBstMidEndSepPunct{\mcitedefaultmidpunct}
{\mcitedefaultendpunct}{\mcitedefaultseppunct}\relax
\EndOfBibitem
\bibitem[Cloke \latin{et~al.}(2015)Cloke, Marangoni, Nguyen, Joshi, Rizzo,
  Bronner, Cao, Louie, Crommie, and Fischer]{cloke_site-specific_2015}
Cloke,~R.~R.; Marangoni,~T.; Nguyen,~G.~D.; Joshi,~T.; Rizzo,~D.~J.;
  Bronner,~C.; Cao,~T.; Louie,~S.~G.; Crommie,~M.~F.; Fischer,~F.~R.
  Site-{Specific} {Substitutional} {Boron} {Doping} of {Semiconducting}
  {Armchair} {Graphene} {Nanoribbons}. \emph{J. Am. Chem. Soc.} \textbf{2015},
  \emph{137}, 8872--8875\relax
\mciteBstWouldAddEndPuncttrue
\mciteSetBstMidEndSepPunct{\mcitedefaultmidpunct}
{\mcitedefaultendpunct}{\mcitedefaultseppunct}\relax
\EndOfBibitem
\bibitem[Carbonell-Sanrom\`a \latin{et~al.}(2017)Carbonell-Sanrom\`a,
  Brandimarte, Balog, Corso, Kawai, Garcia-Lekue, Saito, Yamaguchi, Meyer,
  S\'anchez-Portal, and Pascual]{carbonell-sanroma_quantum_2017}
Carbonell-Sanrom\`a,~E.; Brandimarte,~P.; Balog,~R.; Corso,~M.; Kawai,~S.;
  Garcia-Lekue,~A.; Saito,~S.; Yamaguchi,~S.; Meyer,~E.; S\'anchez-Portal,~D.;
  Pascual,~J.~I. Quantum {Dots} {Embedded} in {Graphene} {Nanoribbons} by
  {Chemical} {Substitution}. \emph{Nano Lett.} \textbf{2017}, \emph{17},
  50--56\relax
\mciteBstWouldAddEndPuncttrue
\mciteSetBstMidEndSepPunct{\mcitedefaultmidpunct}
{\mcitedefaultendpunct}{\mcitedefaultseppunct}\relax
\EndOfBibitem
\bibitem[Carbonell-Sanrom{\`{a}} \latin{et~al.}(2017)Carbonell-Sanrom{\`{a}},
  Hieulle, Vilas-Varela, Brandimarte, Iraola, Barrag{\'{a}}n, Li, Abadia,
  Corso, S{\'{a}}nchez-Portal, Pe{\~{n}}a, and Pascual]{Carbonell2017CN}
Carbonell-Sanrom{\`{a}},~E.; Hieulle,~J.; Vilas-Varela,~M.; Brandimarte,~P.;
  Iraola,~M.; Barrag{\'{a}}n,~A.; Li,~J.; Abadia,~M.; Corso,~M.;
  S{\'{a}}nchez-Portal,~D.; Pe{\~{n}}a,~D.; Pascual,~J.~I. {Doping of Graphene
  Nanoribbons via Functional Group Edge Modification}. \emph{ACS Nano}
  \textbf{2017}, \emph{11}, 7355--7361\relax
\mciteBstWouldAddEndPuncttrue
\mciteSetBstMidEndSepPunct{\mcitedefaultmidpunct}
{\mcitedefaultendpunct}{\mcitedefaultseppunct}\relax
\EndOfBibitem
\bibitem[Durr \latin{et~al.}(2018)Durr, Haberer, Lee, Blackwell, Kalayjian,
  Marangoni, Ihm, Louie, and Fischer]{Durr_JACS2018}
Durr,~R.~A.; Haberer,~D.; Lee,~Y.-L.; Blackwell,~R.; Kalayjian,~A.~M.;
  Marangoni,~T.; Ihm,~J.; Louie,~S.~G.; Fischer,~F.~R. Orbitally Matched
  Edge-Doping in Graphene Nanoribbons. \emph{J. Am. Chem. Soc.} \textbf{2018},
  \emph{140}, 807--813\relax
\mciteBstWouldAddEndPuncttrue
\mciteSetBstMidEndSepPunct{\mcitedefaultmidpunct}
{\mcitedefaultendpunct}{\mcitedefaultseppunct}\relax
\EndOfBibitem
\bibitem[Kawai \latin{et~al.}(2018)Kawai, Nakatsuka, Hatakeyama, Pawlak, Meier,
  Tracey, Meyer, and Foster]{Kawai2018}
Kawai,~S.; Nakatsuka,~S.; Hatakeyama,~T.; Pawlak,~R.; Meier,~T.; Tracey,~J.;
  Meyer,~E.; Foster,~A.~S. {Multiple heteroatom substitution to graphene
  nanoribbon}. \emph{Sci. Adv.} \textbf{2018}, eaar7181\relax
\mciteBstWouldAddEndPuncttrue
\mciteSetBstMidEndSepPunct{\mcitedefaultmidpunct}
{\mcitedefaultendpunct}{\mcitedefaultseppunct}\relax
\EndOfBibitem
\bibitem[Cai \latin{et~al.}(2014)Cai, Pignedoli, Talirz, Ruffieux, S\"ode,
  Liang, Meunier, Berger, Li, Feng, M\"ullen, and Fasel]{cai_graphene_2014}
Cai,~J.; Pignedoli,~C.~A.; Talirz,~L.; Ruffieux,~P.; S\"ode,~H.; Liang,~L.;
  Meunier,~V.; Berger,~R.; Li,~R.; Feng,~X.; M\"ullen,~K.; Fasel,~R. Graphene
  nanoribbon heterojunctions. \emph{Nat. Nanotechnol.} \textbf{2014}, \emph{9},
  896--900\relax
\mciteBstWouldAddEndPuncttrue
\mciteSetBstMidEndSepPunct{\mcitedefaultmidpunct}
{\mcitedefaultendpunct}{\mcitedefaultseppunct}\relax
\EndOfBibitem
\bibitem[Chen \latin{et~al.}(2015)Chen, Cao, Chen, Pedramrazi, Haberer,
  de~Oteyza, Fischer, Louie, and Crommie]{chen_molecular_2015}
Chen,~Y.-C.; Cao,~T.; Chen,~C.; Pedramrazi,~Z.; Haberer,~D.; de~Oteyza,~D.~G.;
  Fischer,~F.~R.; Louie,~S.~G.; Crommie,~M.~F. Molecular bandgap engineering of
  bottom-up synthesized graphene nanoribbon heterojunctions. \emph{Nat.
  Nanotechnol.} \textbf{2015}, \emph{10}, 156--160\relax
\mciteBstWouldAddEndPuncttrue
\mciteSetBstMidEndSepPunct{\mcitedefaultmidpunct}
{\mcitedefaultendpunct}{\mcitedefaultseppunct}\relax
\EndOfBibitem
\bibitem[Li \latin{et~al.}(2018)Li, Merino-D{\'{i}}ez, Carbonell-Sanrom{\`{a}},
  Vilas-Varela, de~Oteyza, Pe{\~{n}}a, Corso, and Pascual]{Li2018}
Li,~J.; Merino-D{\'{i}}ez,~N.; Carbonell-Sanrom{\`{a}},~E.; Vilas-Varela,~M.;
  de~Oteyza,~D.~G.; Pe{\~{n}}a,~D.; Corso,~M.; Pascual,~J.~I. {Survival of spin
  state in magnetic porphyrins contacted by graphene nanoribbons}. \emph{Sci.
  Adv.} \textbf{2018}, \emph{4}, eaaq0582\relax
\mciteBstWouldAddEndPuncttrue
\mciteSetBstMidEndSepPunct{\mcitedefaultmidpunct}
{\mcitedefaultendpunct}{\mcitedefaultseppunct}\relax
\EndOfBibitem
\bibitem[Cao \latin{et~al.}(2017)Cao, Zhao, and Louie]{Cao2017a}
Cao,~T.; Zhao,~F.; Louie,~S.~G. {Topological Phases in Graphene Nanoribbons:
  Junction States, Spin Centers, and Quantum Spin Chains}. \emph{Phys. Rev.
  Lett.} \textbf{2017}, \emph{119}, 076401\relax
\mciteBstWouldAddEndPuncttrue
\mciteSetBstMidEndSepPunct{\mcitedefaultmidpunct}
{\mcitedefaultendpunct}{\mcitedefaultseppunct}\relax
\EndOfBibitem
\bibitem[Merino-D{\'{i}}ez \latin{et~al.}(2018)Merino-D{\'{i}}ez, Li,
  Garcia-Lekue, Vasseur, Vilas-Varela, Carbonell-Sanrom{\`{a}}, Corso, Ortega,
  Pe{\~{n}}a, Pascual, and de~Oteyza]{MerinoJPCL18}
Merino-D{\'{i}}ez,~N.; Li,~J.; Garcia-Lekue,~A.; Vasseur,~G.; Vilas-Varela,~M.;
  Carbonell-Sanrom{\`{a}},~E.; Corso,~M.; Ortega,~J.~E.; Pe{\~{n}}a,~D.;
  Pascual,~J.~I.; de~Oteyza,~D.~G. {Unraveling the Electronic Structure of
  Narrow Atomically Precise Chiral Graphene Nanoribbons}. \emph{J. Phys. Chem.
  Lett.} \textbf{2018}, \emph{9}, 25--30\relax
\mciteBstWouldAddEndPuncttrue
\mciteSetBstMidEndSepPunct{\mcitedefaultmidpunct}
{\mcitedefaultendpunct}{\mcitedefaultseppunct}\relax
\EndOfBibitem
\bibitem[Soler \latin{et~al.}(2002)Soler, Artacho, Gale, Garc\'ia, Junquera,
  Ordej\'on, and S\'anchez-Portal]{Sol02}
Soler,~J.~M.; Artacho,~E.; Gale,~J.~D.; Garc\'ia,~A.; Junquera,~J.;
  Ordej\'on,~P.; S\'anchez-Portal,~D. The SIESTA method for ab initio order- N
  materials simulation. \emph{J. Phys. Condens. Matter} \textbf{2002},
  \emph{14}, 2745\relax
\mciteBstWouldAddEndPuncttrue
\mciteSetBstMidEndSepPunct{\mcitedefaultmidpunct}
{\mcitedefaultendpunct}{\mcitedefaultseppunct}\relax
\EndOfBibitem
\bibitem[Giovannetti \latin{et~al.}(2008)Giovannetti, Khomyakov, Brocks,
  Karpan, van~den Brink, and Kelly]{giovannetti_PRL2008}
Giovannetti,~G.; Khomyakov,~P.~A.; Brocks,~G.; Karpan,~V.~M.; van~den
  Brink,~J.; Kelly,~P.~J. Doping Graphene with Metal Contacts. \emph{Phys. Rev.
  Lett.} \textbf{2008}, \emph{101}, 026803\relax
\mciteBstWouldAddEndPuncttrue
\mciteSetBstMidEndSepPunct{\mcitedefaultmidpunct}
{\mcitedefaultendpunct}{\mcitedefaultseppunct}\relax
\EndOfBibitem
\bibitem[Gonzalez-Lakunza \latin{et~al.}(2008)Gonzalez-Lakunza,
  Fern\'andez-Torrente, Franke, Lorente, Arnau, and
  Pascual]{gonzalez-lakunza_PRL2008}
Gonzalez-Lakunza,~N.; Fern\'andez-Torrente,~I.; Franke,~K.~J.; Lorente,~N.;
  Arnau,~A.; Pascual,~J.~I. Formation of {Dispersive} {Hybrid} {Bands} at an
  {Organic}-{Metal} {Interface}. \emph{Phys. Rev. Lett.} \textbf{2008},
  \emph{100}, 156805\relax
\mciteBstWouldAddEndPuncttrue
\mciteSetBstMidEndSepPunct{\mcitedefaultmidpunct}
{\mcitedefaultendpunct}{\mcitedefaultseppunct}\relax
\EndOfBibitem
\bibitem[Klime\v{s} \latin{et~al.}(2010)Klime\v{s}, Bowler, and
  Michaelides]{klimes_JPCM2010}
Klime\v{s},~J.; Bowler,~D.~R.; Michaelides,~A. Chemical accuracy for the van
  der Waals density functional. \emph{J. Phys. Cond. Matt.} \textbf{2010},
  \emph{22}, 022201\relax
\mciteBstWouldAddEndPuncttrue
\mciteSetBstMidEndSepPunct{\mcitedefaultmidpunct}
{\mcitedefaultendpunct}{\mcitedefaultseppunct}\relax
\EndOfBibitem
\bibitem[Garc\'{\i}a-Gil \latin{et~al.}(2009)Garc\'{\i}a-Gil, Garc\'{\i}a,
  Lorente, and Ordej\'on]{garciagil_PRB2009}
Garc\'{\i}a-Gil,~S.; Garc\'{\i}a,~A.; Lorente,~N.; Ordej\'on,~P. Optimal
  strictly localized basis sets for noble metal surfaces. \emph{Phys. Rev. B}
  \textbf{2009}, \emph{79}, 075441\relax
\mciteBstWouldAddEndPuncttrue
\mciteSetBstMidEndSepPunct{\mcitedefaultmidpunct}
{\mcitedefaultendpunct}{\mcitedefaultseppunct}\relax
\EndOfBibitem
\bibitem[Gross \latin{et~al.}(2011)Gross, Moll, Mohn, Curioni, Meyer, Hanke,
  and Persson]{gross_high-resolution_2011}
Gross,~L.; Moll,~N.; Mohn,~F.; Curioni,~A.; Meyer,~G.; Hanke,~F.; Persson,~M.
  High-{Resolution} {Molecular} {Orbital} {Imaging} {Using} a p -{Wave} {STM}
  {Tip}. \emph{Phys. Rev. Lett.} \textbf{2011}, \emph{107}, 086101\relax
\mciteBstWouldAddEndPuncttrue
\mciteSetBstMidEndSepPunct{\mcitedefaultmidpunct}
{\mcitedefaultendpunct}{\mcitedefaultseppunct}\relax
\EndOfBibitem
\bibitem[Hieulle \latin{et~al.}(2018)Hieulle, Carbonell-Sanrom{\`{a}},
  Vilas-Varela, Garcia-Lekue, Guiti{\'{a}}n, Pe{\~{n}}a, and
  Pascual]{Hieulle2018}
Hieulle,~J.; Carbonell-Sanrom{\`{a}},~E.; Vilas-Varela,~M.; Garcia-Lekue,~A.;
  Guiti{\'{a}}n,~E.; Pe{\~{n}}a,~D.; Pascual,~J.~I. {On-Surface Route for
  Producing Planar Nanographenes with Azulene Moieties}. \emph{Nano Lett.}
  \textbf{2018}, \emph{18}, 418--423\relax
\mciteBstWouldAddEndPuncttrue
\mciteSetBstMidEndSepPunct{\mcitedefaultmidpunct}
{\mcitedefaultendpunct}{\mcitedefaultseppunct}\relax
\EndOfBibitem
\bibitem[Ruffieux \latin{et~al.}(2012)Ruffieux, Cai, Plumb, Patthey, Prezzi,
  Ferretti, Molinari, Feng, M\"ullen, Pignedoli, and
  Fasel]{ruffieux_electronic_2012}
Ruffieux,~P.; Cai,~J.; Plumb,~N.~C.; Patthey,~L.; Prezzi,~D.; Ferretti,~A.;
  Molinari,~E.; Feng,~X.; M\"ullen,~K.; Pignedoli,~C.~A.; Fasel,~R. Electronic
  {Structure} of {Atomically} {Precise} {Graphene} {Nanoribbons}. \emph{ACS
  Nano} \textbf{2012}, \emph{6}, 6930--6935\relax
\mciteBstWouldAddEndPuncttrue
\mciteSetBstMidEndSepPunct{\mcitedefaultmidpunct}
{\mcitedefaultendpunct}{\mcitedefaultseppunct}\relax
\EndOfBibitem
\bibitem[S\"ode \latin{et~al.}(2015)S\"ode, Talirz, Gr\"oning, Pignedoli,
  Berger, Feng, M\"ullen, Fasel, and Ruffieux]{sode_electronic_2015}
S\"ode,~H.; Talirz,~L.; Gr\"oning,~O.; Pignedoli,~C.~A.; Berger,~R.; Feng,~X.;
  M\"ullen,~K.; Fasel,~R.; Ruffieux,~P. Electronic band dispersion of graphene
  nanoribbons via {Fourier}-transformed scanning tunneling spectroscopy.
  \emph{Phys. Rev. B} \textbf{2015}, \emph{91}, 045429\relax
\mciteBstWouldAddEndPuncttrue
\mciteSetBstMidEndSepPunct{\mcitedefaultmidpunct}
{\mcitedefaultendpunct}{\mcitedefaultseppunct}\relax
\EndOfBibitem
\bibitem[Senkovskiy \latin{et~al.}(2018)Senkovskiy, Usachov, Fedorov, Haberer,
  Ehlen, Fischer, and Gr{\"{u}}neis]{Senkovskiy2018}
Senkovskiy,~B.~V.; Usachov,~D.~Y.; Fedorov,~A.~V.; Haberer,~D.; Ehlen,~N.;
  Fischer,~F.~R.; Gr{\"{u}}neis,~A. {Finding the hidden valence band of
  N  =  7 armchair graphene nanoribbons with angle-resolved
  photoemission spectroscopy}. \emph{2D Mater.} \textbf{2018}, \emph{5},
  035007\relax
\mciteBstWouldAddEndPuncttrue
\mciteSetBstMidEndSepPunct{\mcitedefaultmidpunct}
{\mcitedefaultendpunct}{\mcitedefaultseppunct}\relax
\EndOfBibitem
\end{mcitethebibliography}

\end{document}